\documentclass[trackchanges]{aastex7}
\shorttitle{2012 June 3 LPGRE}
\shortauthors{Share et al.}
\begin{document}
%\title{Evidence that SOL2012-06-03 Late Phase $>$100 MeV $\gamma$ Rays are Produced by CME-Driven Shock Accelerated Suprathermal Ions from the Flare}
\title{Evidence that SOL2012-06-03 Late Phase $\gamma$ Rays are Produced by $>$300 MeV Protons from CME-Shock Acceleration of Suprathermals from the Flare}  

\author[0000-0002-9204-3934]{Gerald H. Share}
\affiliation{ Astronomy Department, University of Maryland, College Park, MD 20740, USA}
%\affiliation{ TSC, Resident at the Naval Research Laboratory, Washington, DC 20375-5352, USA}
\email[show]{gershare@aol.com}\
\author{Ronald J. Murphy}
\affiliation{Retired}
\email{ronald.murphy@hotmail.com}

\begin{abstract}

A recent paper on SOL2012-06-03 reported the detection for the first time of two distinct phases of $>$100 MeV $\gamma$-radiation indicating separate acceleration processes.  But such two-phase emission has been seen before and was first observed in SOL1982-06-03.  The second phase is known as Late Phase Gamma-Ray Emission (LPGRE) and was cataloged for $>$40 solar eruptions, including SOL2012-06-03.  Here we provide evidence that the second SOL2012-06-03 $\pi$-decay peak is the onset of LPGRE that lasted for $>$8 min. Its delay from the impulsive X-ray peak is consistent with the time it would take flare-produced suprathermal protons to overtake the expanding CME and be accelerated by its shock.   The high accelerated ion-to-electron ratio in SOL2012-06-03 and other LPGRE events is consistent with the ratio observed in gradual SEP events produced by shocks and is inconsistent with ratios typically found in impulsive flares and solar energetic particle events produced by reconnection.

 \end{abstract}

%\keywords{Sun: corona --- Sun: chromosphere --- Sun: flares --- Sun: electrons --- Sun: X-rays, gamma rays}


%\section{A short history of AASTeX} 

\section{Introduction} 

\citet{pesc25} recently reported that SOL2012-06-03 ``represents the first time that two distinct phases are observed for $>$100 MeV emission."  Here, however, we argue that the emission has the characteristics of what has been called Late Phase Gamma-Ray Emission (LPGRE), first observed in 1982.  We summarize the characteristics of LPGRE in $\S$\ref{sec:lpgre} and summarize the observations of SOL2012-06-03 that justify our different interpretation of this eruption in $\S$\ref{sec:0603}. 

\section{Characteristics of Late Phase Gamma-Ray Emission} \label{sec:lpgre}
\begin{figure}[h!]
\gridline{\fig{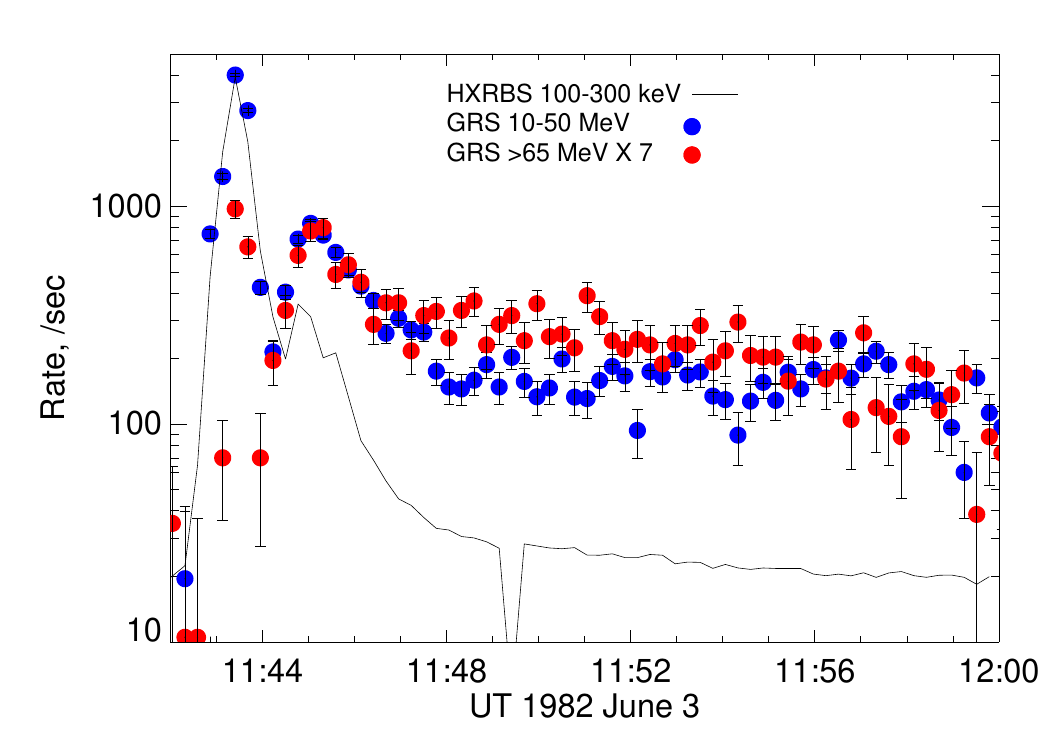}{0.5\textwidth}{(a)}
          \fig{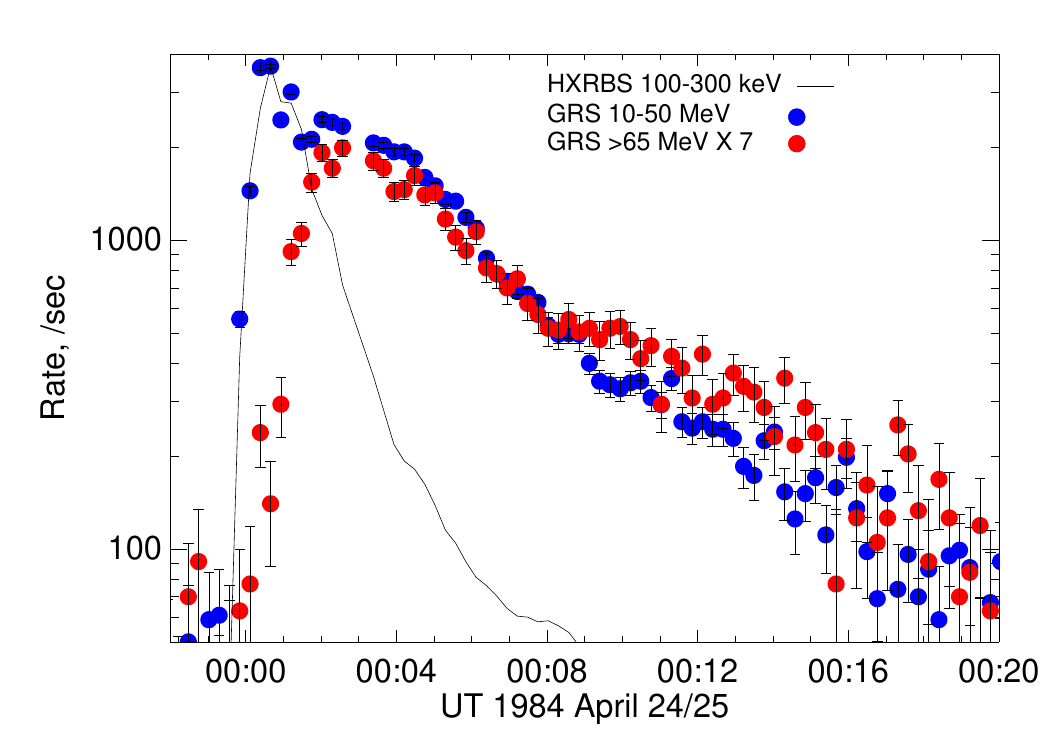}{0.5\textwidth}{(b)}}
\gridline{\fig{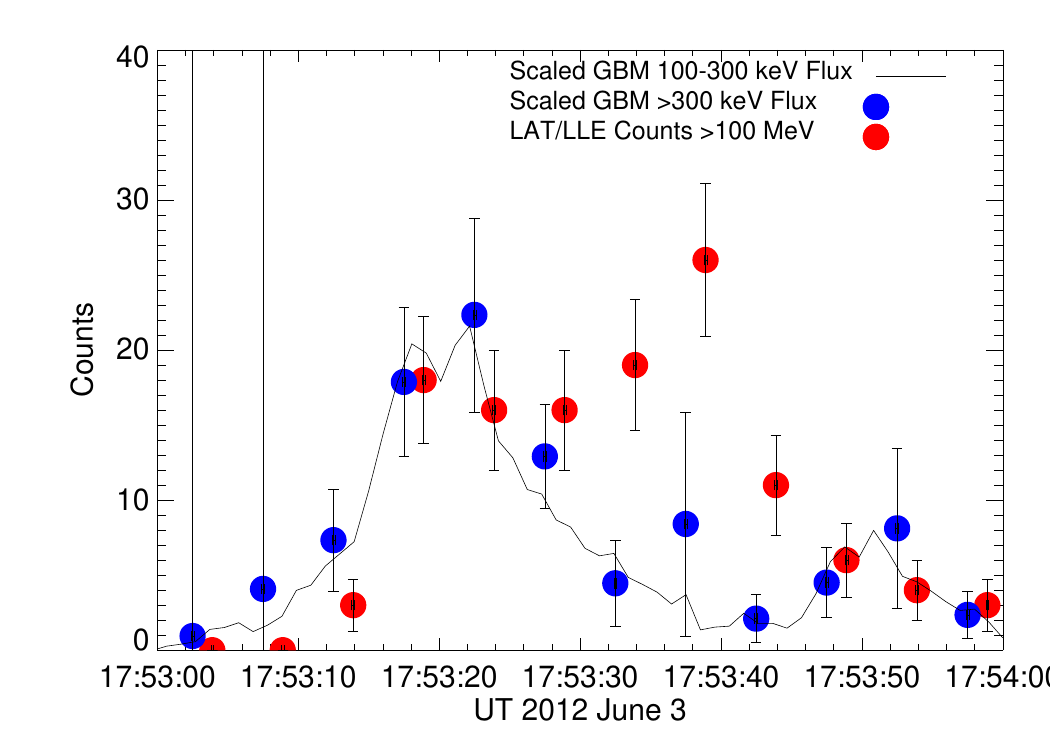}{0.5\textwidth}{(c)}}       
\caption{Three solar eruptions revealing two distinct phases of high-energy emission with the second dominated by protons producing $\pi$-decay emission (GRS 10--50~MeV/$>$65~MeV flux ratios $\gtrapprox$7). The 100--300 keV rates define the impulsive phase.}
\label{thsmm}
\end{figure}

Two distinct phases of high-energy $\gamma$-ray emission were first observed by the {\it Solar Maximum Mission} Gamma-Ray Spectrometer ({\it SMM}/GRS) in SOL1982-06-03 \citep{forr86}.  The {\it SMM} Hard X-Ray Burst Spectrometer (HXRBS) 100--300~keV hard X-ray rates and the GRS 10--50~MeV and $>$65~MeV $\gamma$-ray rates plotted in Figure~\ref{thsmm}(a) peaked during the impulsive phase of the flare.  The second phase began within a minute of that peak and persisted for at least tens of minutes. The (10--50~MeV)/($>$65~MeV) flux ratio of $\gtrapprox$7 during this phase is what was expected if the spectrum was dominated by $\pi$-decay emission.  We plot the time history of a similar event, SOL1984-04-25, in Figure~\ref{thsmm}(b), in which no impulsive $>$65 MeV $\gamma$-rays were observed.  \citet{ryan00} summarized the characteristics of 13 similar events from 1982 to 1991 and called the second phase emission `long duration gamma-ray flares' and defined them as ``time periods of a fraction of an hour to hours after the impulsive phase while other common flare emissions (e.g., X-rays) are absent or greatly diminished." The hours-long events are often referred to as `sustained gamma-ray emission' because their temporal characteristics were distinctly different from the associated impulsive flares.  Because the second phase of some events began within a minute of the impulsive flare and had only minutes-long durations (e.g. SOL2011-08-09, SOL2011-09-06, SOL2011-09-24, and SOL2012-06-03), \citet{shar18} adopted the name `Late Phase Gamma-Ray Emission' (LPGRE).

There is a growing consensus that LPGRE is due to protons accelerated by fast CME-driven shocks onto field lines returning to the Sun (e.g. \citet{plot17,jin18,koul20,gopa21,pesc22}).  The fact that onsets of $>$100 MeV emission occurred within a minute of the impulsive peak is consistent with studies indicating that CME shocks can form at heights as low as 1.2 R$_s$ \citep{gopa13}.  \citet{shar18} suggested that CME-driven shock acceleration of suprathermal ions from the accompanying flare could provide significant contributions to both the LPGRE and gradual SEP fluxes.  They provided evidence that the onset delay of 18 LPGRE events could be explained by the time it would take flare-produced suprathermal protons to overtake the CME-driven shock and be accelerated onto magnetic field lines returning to the Sun.  A low electron-to-proton ratio is characteristic of both LPGRE and gradual SEP events, suggesting that they have a common origin in CME shock acceleration. \citet{cliv16} found that the (0.3 -- 0.7 MeV electron)/($>$100 MeV proton) flux ratio in gradual events has a value of $\sim500$ which is over two orders of magnitude smaller than the ratio in impulsive SEP events. This ratio can be estimated in the behind-the-limb SOL2014-09-01 eruption because both late phase hard X-ray and $>$100 MeV $\gamma$-ray emission were observed. We estimate that during the late phase between 11:06--11:20 UT there were $\sim$ 3 $\times$ 10$^{31}$ protons with energies $>$100 MeV and $\sim$1.5 $\times$ 10$^{34}$ electrons between 300 and 700 keV.  This electron-to-proton ratio of $\sim$500 is consistent with gradual SEP events, suggesting that CME shock acceleration is responsible for LPGRE. Only upper limits on the ratio can be obtained for the other eruptions observed up until 2015 July.

\section{SOL2012-06-03 Second Peak is the Onset of LPGRE}\label{sec:0603}

The LPGRE characteristics of SOL2012-06-03 were detailed in Appendix~C16 of \citet{shar18}. The event began with an $\sim$1~min impulsive phase with two $>$100 MeV peaks followed by decreasing emission that persisted for 8~min until the telescope pointed away from the Sun. The decay-phase characteristics were consistent with those of LPGRE: 1) spectrum consistent with the decay of pions produced by protons having a power-law index of 4.4 $\pm$ 0.5 above 300 MeV and a flatter spectrum at lower energies; 2) the late/impulsive phase flux ratio was $\sim$4 if both peaks are assumed to be from the flare; 3) $>$100 keV X-ray, Type II and III radio emissions, and a CME speed of $\sim$900 km s$^{-1}$.   In Figure~\ref{thsmm}(c) we plot $>$100~MeV  counts in 5~s intervals. The first peak was coincident with the peak in the {\it Fermi} Gamma-Ray Burst Monitor (GBM)  NaI 100--300~keV flux. As discussed by \citet{pesc25}, the second $>$100~MeV peak\footnote{The significance of the two peaks claimed by \citet{pesc25} is not evident in this linear plot because the uncertainties in the logarithmic plots of that paper were incorrectly displayed. An F-test of the $\chi^2$ analysis in \citet{pesc25} only indicates a 90\% probability that two Gaussians fit the data better than does a single one.} occurred while there was no peak in the decaying X-ray flux.  As this is the basic characteristic of LPGRE, it is likely that the second peak was the start of $>$100 MeV LPGRE.  There is additional weak evidence for such an association, as our spectral fits indicate that the power-law index hardened from 5.6 $\pm$ 1.4 during the first peak to 4.3 $\pm$ 1.0 during the second peak, a value consistent what was measured in the LPGRE decay phase \citep{shar18,pesc25}. The apparently harder spectrum of the second peak agrees with the data in Table 2 of \citet{pesc25}. If the second peak was the onset of LPGRE, then the late/impulsive phase flux ratio increases to $\sim$15 .

The $\sim$17 s delay of the second peak in SOL2012-06-03 can be explained by the acceleration of flare suprathermal ions by a CME-driven shock as found in other LPGRE events. For a 900 km s$^{-1}$ CME with 17:53:00 UT onset, 75 keV suprathermal protons released at the peak of the hard X-ray emission, would overtake the CME-driven shock in about 5 s.  This is consistent with the peak time of $>$100 MeV $\gamma$ rays, assuming that the protons can be accelerated to such high energies in $\sim$10 s and promptly return to the Sun.

\section{Summary}

In this note we provide evidence that the second peak of $>$100 MeV $\gamma$ radiation in SOL2012-06-03 was in fact the onset of Late Phase Gamma Ray Emission (LPGRE), as detailed for that eruption and 29 others by \citet{shar18}.

\bibliography{references}{}
\bibliographystyle{aasjournal}

\end{document}